\documentclass[times]{smrauth}

\usepackage{moreverb}
\usepackage{subfigure}
\usepackage{caption}
\usepackage{graphicx}
\usepackage{cite}

\usepackage[colorlinks,bookmarksopen,bookmarksnumbered,citecolor=red,urlcolor=red]{hyperref}

\newcommand\BibTeX{{\rmfamily B\kern-.05em \textsc{i\kern-.025em b}\kern-.08em
T\kern-.1667em\lower.7ex\hbox{E}\kern-.125emX}}

\usepackage{wasysym}

\begin{document}

\runningheads{M. Kuhrmann, D. M\'{e}ndez Fern\'{a}ndez, T. Ternit\'{e}}{On the Use of Variability Operations in the V-Modell XT Software Process Line}

\title{On the Use of Variability Operations in the V-Modell XT Software Process Line}

\author{Marco Kuhrmann\affil{1}\corrauth, Daniel M\'{e}ndez Fern\'{a}ndez\affil{2}, Thomas Ternit\'{e}\affil{3}} 

\address{
	\affilnum{1}{University of Southern Denmark, The M\ae rsk Mc-Kinney M\o ller Institute \& The Center for Energy Informatics, Odense, Denmark}\break
	\affilnum{2}{Technische Universit\"at M\"unchen, Faculty of Informatics, Garching, Germany}\break
	\affilnum{3}{Technische Universit\"at Clausthal, Department of Informatics, Clausthal-Zellerfeld, Germany}
}

\corraddr{University of Southern Denmark, The M\ae rsk Mc-Kinney M\o ller Institute \& The Center for Energy Informatics, Campusvej 55, 5230 Odense M, Denmark. E-Mail: kuhrmann@mmmi.sdu.dk}

\begin{abstract}
Software process lines provide a systematic approach to develop and manage software processes. A process line defines a reference process containing general process assets, whereas a well-defined customization approach allows process engineers to create new process variants, e.g., by extending or modifying process assets. Variability operations are an instrument to realize flexibility in the V-Modell~XT process line by explicitly declaring required modifications, which are applied in a later step to create a procedurally generated company-specific process. However, little is yet known about which variability operations are suitable in practice. In this article, we present a study on the feasibility of variability operations to support the development of software process lines in the context of the German V-Modell~XT.  We analyze which variability operations are defined and practically used, and if not used, why. We provide an initial catalog of variability operations as an improvement proposal for other process models. Our findings show that 69 variability operation types are defined across several metamodel versions of which, however, 25 remain unused. The found variability operations allow for systematically modifying the content of process model elements and the process documentation, and they allow for altering the structure of a process model and its description. Furthermore, we also find that variability operations can help process engineers to compensate process metamodel evolution.
\end{abstract}


\maketitle

\section{Introduction}
\label{sec:Introduction}
The V-Modell is the standard software process model for public sector IT development projects in Germany. It has a long history beginning with its first release in 1992, and it was improved in several iterations. Since 2005, the revised version of the V-Modell~XT became subject to numerous adaptations, which were, initially, conducted following a ``copy \& change'' procedure in which company-specific process variants were realized by directly modifying a local copy of the reference process. As the reference process evolved \cite{kms2013}, this approach caused serious problems, e.g., when integrating updated contents, figuring out what particular customizations were affected by newer reference contents, and when migrating existing content to a new process metamodel. Much effort has been spent to analyze the evolved variants (see, e.g.,~\cite{OM09,OMR09,OS07}). However, only the changes could be analyzed and documented. Efficiently integrating evolved model contents with customized ones to create a new version of the company-specific process remained a critical and unresolved task.

In response to this problem, in the WEIT\footnote{\textbf{WE}iterentwicklung des \textbf{IT}-Entwicklungsstandards des Bundes---SPI project to improve the standard IT development process model of the German government.} project, it was decided to adopt principles from \emph{Software Process Lines} (SPrL, \cite{rombach2005})  to maintain the reference process \emph{and} its variants to allow for evolution and (automatic) updates. Special attention was paid to the customization\footnote{The V-Modell XT supports process customization at different levels: At the organization level, companies can develop a company-specific software process variant, which can be grounded in a reference model. Furthermore, the V-Modell~XT supports the project-specific tailoring in which a (company-specific) process is tailored according to the actual project context. This article addresses the organization level and presents an instrument used to develop a process variant from a reference process in the context of a process line. Referring to the nomenclature from \cite{HolisticPMELifecycle14} (PD: Process Description, MM: Metamodel), this article covers the areas PDMM, PD, Adapt PDMM, and Create/Update PD. The article addresses process engineers (PE) rather than project managers.} approach to support process engineers in, among other things, using \emph{typed variability operations} \cite{tt2009} as a declarative instrument to systematically derive a company-specific process variant from a reference model while ensuring consistency and addressing compliance requirements.

\paragraph{Problem Statement}
\label{sec:ProblemStatement}
While defining the variability operations for the V-Modell~XT, a major problem was to define a set of suitable variability operations. Available approaches are either of conceptual nature \cite{nz12} or they focus on general concepts \cite{OMG2005} and require further refinement. That is, process engineers need to develop their own portfolio of variability instruments negatively impacting the SPrL, e.g., due to incompatible sets of variability operations, and potential losses of compliance. Missing is a set of actionable variability operations to support process variant development within a comprehensive SPrL. Furthermore, we still lack long-term studies analyzing the feasibility of SPrL approaches \cite{al.:2020ly,Carvalho:2014zr}.

\paragraph{Contribution}
\label{sec:Contribution}
In this article, we contribute an exploratory study on the application of variability operations to realize a comprehensive SPrL.  We study a snapshot from 2013 of the V-Modell~XT process line \cite{kms2013} in which we investigated the V-Modell~XT reference model and 5 of its variants using the built-in SPrL features. 
We contribute an initial catalog of variability operations as implemented in the V-Modell~XT \cite{Kuhrmann:2014fk}, and analyze the feasibility of this instrument. We investigate which variability operations were defined and to what extent those were used in practice. Furthermore, we analyze settings in which variability operations were not used and provide a rationale. In addition to our previously published study \cite{Kuhrmann:2014:RSP:2600821.2600833}, we provide a broader perspective on the context of our contribution and provide more details on the variability operation instrument.

\paragraph{Outline}
\label{sec:Outline}
The remainder of this article is organized as follows: Section~\ref{sec:ContextAndTerminology} introduces the basic concepts and terminology used and summarizes the related work. In Section~\ref{sec:StudyDesign}, we present the study design including our research questions, the case selection, and the data collection and analysis procedures. In Section~\ref{sec:StudyResults}, we finally present our findings, before giving a conclusion in Section~\ref{sec:Conc}.


\section{Fundamentals \& Related Work}
\label{sec:RelatedWork}
In this section, we provide a brief overview of the key concepts behind the V-Modell~XT process line, and we discuss related work.

\paragraph{Context \& Used Terminology}
\label{sec:ContextAndTerminology}
We analyzed the V-Modell~XT SPrL \cite{kms2013}, which is a process framework (including, e.g., metamodels, tools, reference implementations, and guidelines) with built-in SPrL features. In order to provide understanding of variability operations as one of these features, in Table~\ref{tab:VMXT:ConceptsAndTerminology}, we introduce the basic concepts and underlying terminology.

\begin{table}[htbp]
\renewcommand{\arraystretch}{1.3}
\footnotesize

\caption{Basic terminology and concepts in the V-Modell XT process line.} 
\label{tab:VMXT:ConceptsAndTerminology}
\label{sec:VMXTProcessLine}
\centering
\begin{tabular}{p{0.12\textwidth}p{0.82\textwidth}}
\hline
	\textbf{Concept} & \textbf{Description} \\
\hline
	Model,\newline Metamodel\newline and Modules & The V-Modell~XT is a modular, metamodel-based framework to define software processes. The metamodel~\cite{TK09} defines the \emph{process language} to create single processes and SPrLs. A \emph{V-Modell-variant} logically consist of two models: (1) a structure model contains all (atomic) model elements, and (2) an overlaying dependency model connects all model elements. Hence, if dependencies are contained in \emph{process modules}, the configuration of such modules directly influences the dependency model, which allows for a comprehensive tailoring \cite{Kuhrmann2008}. During customization, all elements of these packages can be extended, altered, and so on. \\
	Process\newline Variants & The metamodel supports hierarchically organized process variants---even the reference model is seen as a variant \cite{tt2009,ternite2010}. Creating a new variant requires a \emph{reference model} on which the variant is based. A variant can be regarded as an \emph{extension} applied to a reference model. Since all V-Modell-variants are (or should be) based on the same metamodel, each variant may contain a complete process description. As all model elements from the reference model are accessible from a variant, a variant can refer to and, thus, integrate and modify any reference model element. A merge tool creates an integrated process description from the variants. New assets introduced by the variant will then be integrated with the reference model, exclusions will be deleted, and variability operations will be processed. \\
	V-Modell~XT\newline Process Line & The V-Modell~XT family consists of a reference process and a number of derived variants \cite{kms2013}. The snapshot, which is the case for our study (Figure~\ref{fig:VMProcessLineOverview}), shows the reference model and two kinds of variants: We find variants created as a direct modification of a local copy (``old'' scheme), and we find 5 variants using the built-in SPrL features. Variants using the SPrL features can reuse content from the reference model and support automatic updates. If a new version of the reference process is released, in the simple case, the merge tool automatically updates a variant, e.g., by computing the variability operations again. \\
\hline
\end{tabular}
\end{table}

\begin{figure*}[ht]
	\begin{center}
		\includegraphics[width=\textwidth]{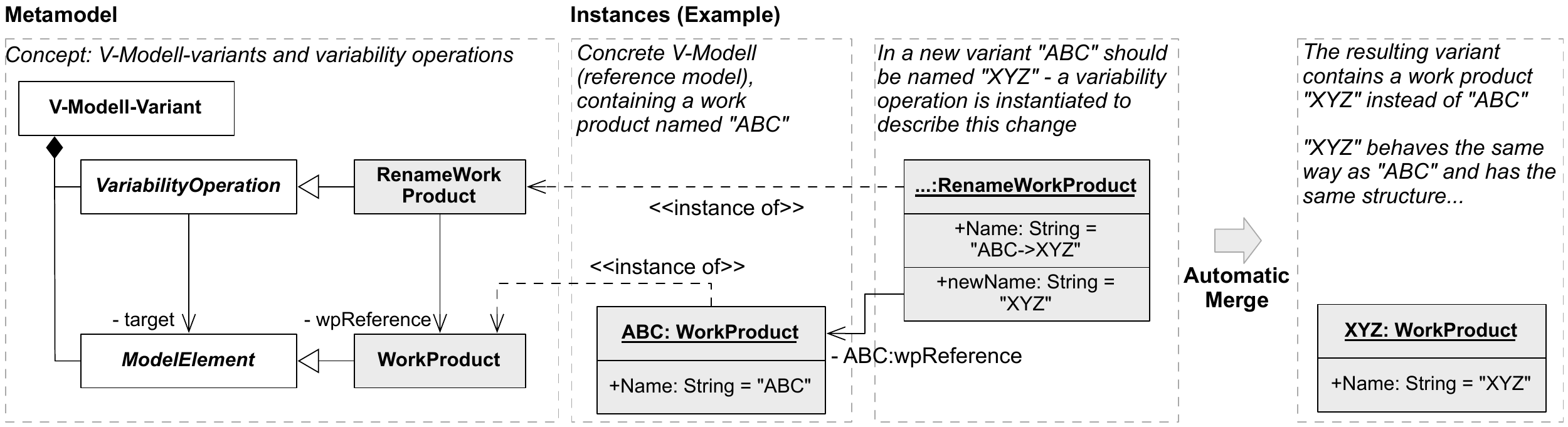}
	\caption{Variability operations (concept and example).}
	\label{fig:VarOpsConcept}
	\end{center}
\end{figure*}

\paragraph{Variability Operations}
As part of the customization framework, variability operations allow a process variant to modify contents and structure of the reference model \cite{tt2009}. A variability operation is a model element that declares a change, e.g., renaming of elements, adding description text, or restructuring dependencies (Figure~\ref{fig:VarOpsConcept}). Furthermore, the instrument described in this article provides \emph{typed} variability operations, i.e., the operations have a specific purpose. For example, if the task is to rename a role, the respective operation is named \emph{RenameRole} and it only refers to elements of type \emph{Role} in the process model. The complex variability operations---as provided to process engineers---are realized by assembling certain atomic model-transformation operations, such as \emph{RenameElement}, \emph{AddText}, \emph{ReplaceText}, or \emph{SwapRefences} \cite{ternite2010}. From these atomic operations, if applicable, process engineers may also compose new operation types to extend the set of available operations in response to particular customer requirements (Section~\ref{sec:RQ1}).

A particular variability operation exemplar refers to a model element in the reference process, and describes how the referred element will be treated during the merge procedure in which the reference model and an extension model are processed in order to compile the company-specific process variant (Table~\ref{tab:VMXT:ConceptsAndTerminology}), i.e., variability operations are used during the definition of a company-specific process. Utilizing variability operations ensures that the source files of the reference model are not touched, as all change declarations are contained in an externally managed extension model. During the merge procedure, the  information provided by variability operations is operationalized by the merge tool. For example, the declaration of renaming a work product has to be evaluated and executed during that merge. If an extension model contains an exemplar of the \emph{RenameWorkProduct} operation, this has to be interpreted as a rename operation (execute the atomic operation \emph{RenameElement}) on the referred instance of `WorkProduct' during the merge (Figure~\ref{fig:VarOpsConcept}). After the merge procedure, the integrated company-specific process variant can be deployed to the projects in which the process is subject to further tailoring and, eventually, application.

\paragraph{Related Work}
In~\cite{rombach2005}, Rombach votes for organizing comprehensive software processes similar to product lines. To this end, SPrLs consist of a stable core (commonalities) and variable parts (variabilities)~\cite{cdkt2001,cohen1999,nz12, seipl}. Software process lines propose advantages regarding the organization and management of process knowledge and the systematic creation of reusable process assets to ease process variant development. A software process line is a framework for a directed and proactive process construction and management. Yet only few publications deal with self-contained approaches providing support for process engineers. For instance, in \cite{Bastarrica:1900dq}, Alegr\'{i}a and Bastarrica present CASPER, which they define as meta-process to support the construction of project-specific processes. Mart\'{i}nez-Ruiz et al.\ \cite{al.:1900bh} and Oliveira et al.\ \cite{Oliveira:1900cr} propose extensions of SPEM's  variability constructors, and address the problem that SPEM, basically, provides generic variability operations and modularization concepts, but does not provide explicit and context-specific variability constructors. 
SPEM \cite{OMG2005} and the V-Modell~XT \cite{TK09} explicitly define variability operations. Process assets built on these metamodels can extend or modify other process assets, and they can be configured from certain (process) modules.
However, in contrast to SPEM that only provides a generic concept of variability, the V-Modell~XT explicitly defines a process variant concept and provides an extensive set of \emph{typed} variability operations for fine-grained model manipulations. In \cite{ternite2010}, Ternit\'{e} describes typed variability operations to support software process variability (Figure~\ref{fig:VarOpsConcept}). This instrument is used in the V-Modell XT to realize a complex software process line (Figure~\ref{fig:VMProcessLineOverview}).

Instead of constructively defining process variants, M\"unch et al.\ focus on an evolutionary approach that comprises: (1) scoping processes to work out where variability is needed~\cite{AKM+08,AKM+09} by determining the properties an actual process has and by identifying commonalities/pattern to infer needed variability; (2) providing rationale during process evolution~\cite{OM09,OMR09,OS07}; and (3) analyzing differences of evolved model variants. This approach is based on an \emph{a posteriori} observation of the evolved subject. 

Although considered of particular interest, SPrL concepts are still considered immature due to a lack of empirical evidence for their feasibility \cite{al.:2020ly,Carvalho:2014zr}, or due to the absence of a meaningful notation for variable processes \cite{Martinez-Ruiz:2012fk}. Also, a deeper understanding of the variability instruments is in general yet missing. With the study at hand, we close this gap in literature.

\section{Study Design}
\label{sec:StudyDesign}
In this section, we present the study design. After defining the goal and the research questions, we describe how we selected the case. Finally, we describe how we collected and analyzed the data.

\subsection{Research Questions}

Our overall objective is to study the feasibility and the practical application of variability operations to support the (long-term) development and the maintenance of SPrLs. To this end, we investigate which variability operations are implemented in general and to what extend these operations are used. In a second step, we study settings in which variability operations were not used and why. We define our research questions in Table~\ref{tab:ResearchQuestions}.

\begin{table}[ht]
\renewcommand{\arraystretch}{1.3}
\footnotesize

\caption{Research questions.} 
\label{tab:ResearchQuestions}
\centering
\begin{tabular}{lp{0.9\textwidth}}
\hline
	\textbf{No.} & \textbf{Description and Rationale} \\
\hline
\hline
	$\text{RQ}_{1}$ & \emph{Which variability operations are defined to realize the process line?}\newline Since most related work discusses---if at all---variability operations in a generic manner, our first research question aims to identify a set of variability operations to create a catalog. \\
	$\text{RQ}_{2}$ & \emph{Which variability operations are practically used to which extent?}\newline This question aims to investigate the feasibility of the found variability operations. We analyzed to which extent the found variability operation types were actually used in particular process variants. \\
	$\text{RQ}_{3}$ & \emph{In which settings are variability operations not used and why?}\newline We aim to investigate settings that are potentially inappropriate for variability operations. We studied settings in which variability operations were not used, and we investigated the respective settings, analyzed the instruments used instead of variability operations, and provide a rationale. \\
\hline
\end{tabular}
\end{table}

\subsection{Case Selection}
\label{sec:CaseAndSubjectSelection}
We opted for the V-Modell~XT to collect and analyze variability operations. As we were interested in the variability operations and their use, we only considered such variants using the built-in SPrL features---all other variants are out of scope.

\subsection{Data Collection Procedure}
\label{sec:DataCollectionProc}
To answer $\text{RQ}_{1}$ and $\text{RQ}_{2}$, we used a tool to export lists of the variability operations defined and used. All information was collected by parsing the models' XML files and storing the data in a spreadsheet. 
Therefore, we first analyzed the respective metamodel on which a process variant is based and gathered all defined \emph{variability operation types}. In a second step, we exported the \emph{variability operation exemplars} as defined in the process models (in this step, we also analyzed which version of the metamodel defines an operation to track the metamodel evolution). We repeated the export process for all considered variants to create (1) a consolidated list of operation types across all versions of the metamodel, (2) process-variant-specific lists of variability operation exemplars, and (3) an aggregated list of all variability operations, their type, number of exemplars, and so forth. In order to answer $\text{RQ}_{3}$, we had to (manually) inspect the considered process variants. We compared the merged process definition with its sources for added, modified, and/or removed process assets that are not defined using variability operations. The outcome of this investigation was also stored in a spreadsheet.

\subsection{Analysis Procedure}
\label{sec:AnalysisProc}
Due to the low number of cases, we present the results as data tables and simple charts, and qualitatively analyze and interpret the results.


\section{Study Results}
\label{sec:StudyResults}
We first give a description of the case, before summarizing the results structured according to the research questions.

\subsection{Case Description}
\label{sec:CaseAndSubjectDescription}
\begin{figure*}[t]
	\begin{center}
		\includegraphics[width=\textwidth]{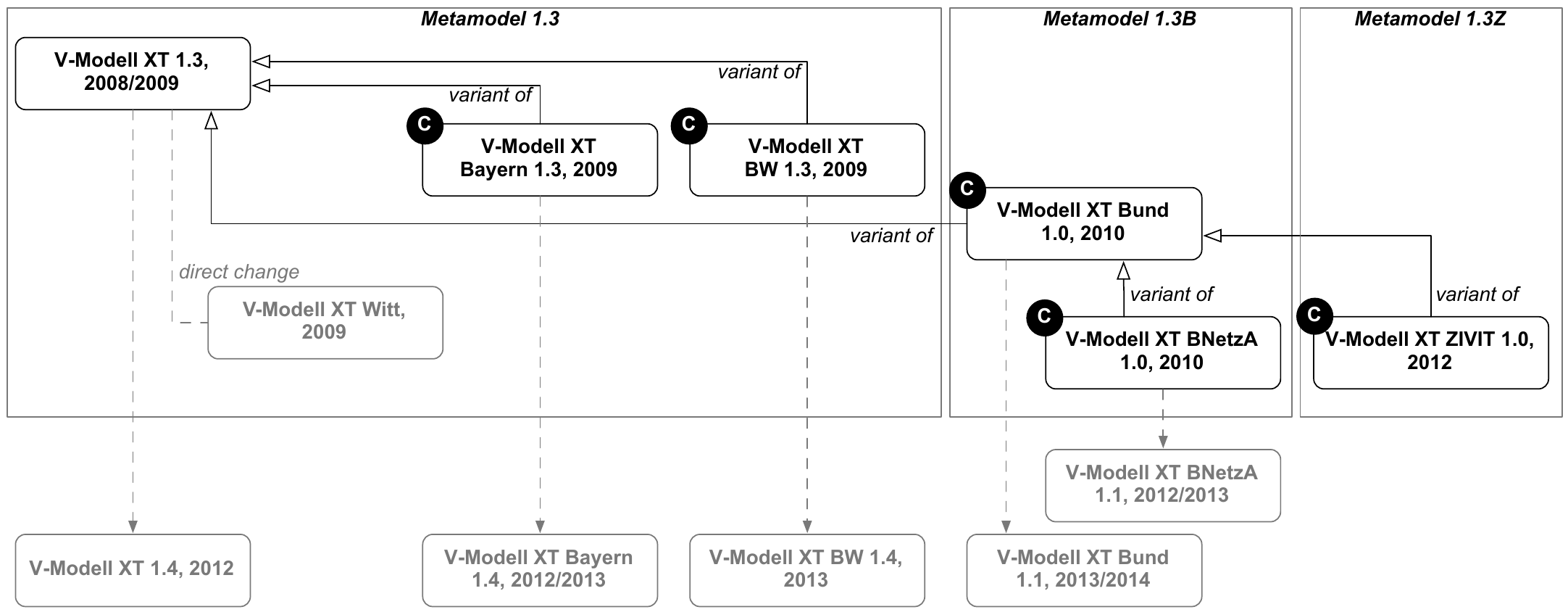}
	\caption{Snapshot of the V-Modell~XT software process line (variants marked with ``C'' form our case).}
	\label{fig:VMProcessLineOverview}
	\end{center}
\end{figure*}

As case, we opt for the V-Modell~XT and the set of 5 variants that use the built-in SPrL features. Figure~\ref{fig:VMProcessLineOverview} shows a snapshot; the highlighted variants are subject to the study---the other variants do not use the SPrL features and, thus, are out of scope.  
Our study is based on the V-Modell~XT~1.3\footnote{Although the V-Modell~XT~1.4 was released in 2012, no variants using this version as reference model were available when we conducted the study.} and we refer to this version as the \emph{reference model} on which, finally, all variants\footnote{\textbf{Note}: Except for the variants \emph{V-Modell~XT~1.3} and \emph{V-Modell~XT Bund~1.0}, for confidentiality reasons, we are not allowed to relate the findings to a variant from Figure~\ref{fig:VMProcessLineOverview}; we provide the data, but anonymized. A collection of publicly available material is provided by the Weit e.V.: \url{www.weit-verein.de/varianten.html} (in German).} are built. 
Each variant, except for the V-Modell~XT~1.3, points to its parent (Figure~\ref{fig:VMProcessLineOverview} also shows that the SPrL builds a ``family tree'' in which a derived variant can be a reference model for further variants).

\subsection{RQ~1: Defined Variability Operation Types}
\label{sec:RQ1}
Since the variants under consideration use different versions/variants of the metamodel (Figure~\ref{fig:VMProcessLineOverview}), we first analyze (1) which variability operation types are defined and (2) which metamodel defines a particular operation type. 

Due to space limitations, the complete list (the catalog) of variability operations is not part of this article, but is available for download \cite{Kuhrmann:2014fk}. In total, the V-Modell~XT metamodel provides process engineers with 69 variability operation types, which are defined by the metamodel versions ``1.3'' and ``1.3B'' (the metamodel ``1.3Z'' does not define new operation types). Figure~\ref{fig:VarOpsComplete} summarizes the number of defined variability operation types per operation group\footnote{An operation group comprises all operation types that are logically related (e.g., changes on work products).} and per defining metamodel version. Furthermore, Figure~\ref{fig:VarOpsComplete} also reflects the evolution of the metamodel---35 new operation types were introduced in the metamodel ``1.3B'' (two years after the publication of the reference model 1.3, which defines 34 variability operation types).

\paragraph{Interpretation}
We found variability operation types defined in two metamodel versions. Moreover, the number of operation types doubled. An explanation can be found in the metamodel's evolution. The metamodel ``1.3B'' got a substantial improvement, which was based on customer requirements, whereas the initial set of operation types was derived from known improvements at this time and compliance requirements in the context of a certification program. So far, the growing number of operation types indicates that the mechanism ``variability operation'' can be used to foster flexibility in a process line (in response to customer requirements).

\subsection{RQ~2: Variability Operation Usage}
\label{sec:RQ2:Use}
The second research question aims at investigating which of the defined operation types are used in practice.  
\begin{table}[t]
\renewcommand{\arraystretch}{1.15}
\footnotesize

\caption{Used variability operations by variant, operation type, and metamodel release.} 
\label{tab:VMXT:ExemplarsByOperationGroup}
\label{tab:VarOpsByTypeMMRelease}
\centering
\begin{tabular}{l|cc|c|cc|c|cc|c|cc|c}
\hline
\textbf{Operation Group} & \multicolumn{3}{|c}{\textbf{Variant Bund}} & \multicolumn{3}{|c}{\textbf{Variant A}} & \multicolumn{3}{|c}{\textbf{Variant B}} & \multicolumn{3}{|c}{\textbf{Variant C}} \\
	                & \textbf{1.3}  & \textbf{1.3B} & \textbf{$\sum$} & \textbf{1.3} & \textbf{1.3B} & \textbf{$\sum$} & \textbf{1.3}  & \textbf{1.3B} & \textbf{$\sum$} & \textbf{1.3}  & \textbf{1.3B} & \textbf{$\sum$}  \\
\hline
Discipline Variations            & 1	& 12	& 13	        & 0	    & 0	      & 0	& 0	        & 0	         & 0	& 2	       & 1	        & 3 \\ 
Work Product Variations          & 1	& 11	& 12	        & 3	    & 0	      & 3	& 3	        & 2	         & 5	& 0	       & 1	        & 1 \\
Topic Variations                 & 2	& 19	& 21	        & 5	    & 0	      & 5	& 9	        & 0	         & 9	& 1	       & 0	        & 1 \\
Activity Variations              & 0	& 1	    & 1	            & 1	    & 0	      & 1	& 0	        & 0	         & 0	& 0	       & 0	        & 0 \\
Task Variations                  & 0	& 0	    & 0	            & 0	    & 0	      & 0	& 0	        & 0	         & 0	& 0	       & 0	        & 0 \\
Role Variations                  & 4	& 47	& 51	        & 0	    & 0	      & 0	& 24	    & 17	     & 41	& 14	   & 18	        & 32 \\
Tailoring Variations             & 0	& 4	    & 4	            & 1	    & 0	      & 1	& 1	        & 0	         & 1	& 0	       & 0	        & 0 \\
Decision Gate Variations         & 0	& 10	& 10	        & 3	    & 0	      & 3	& 1	        & 0	         & 1	& 0	       & 0	        & 0 \\
Description Replacements         & 25	& 0	    & 25	        & 1	    & 0	      & 1	& 4	        & 0	         & 4	& 24	   & 0	        & 24 \\
Description Add-ons              & 0	& 0	    & 0	            & 2	    & 0	      & 2	& 4	        & 0	         & 4	& 1	       & 0	        & 1 \\
Description Re-Arragements       & 1	& 1	    & 2	            & 1	    & 0	      & 1	& 6	        & 0	         & 6	& 5	       & 0	        & 5 \\
Description Removements          & 0	& 3	    & 3	            & 0	    & 0	      & 0	& 0	        & 1	         & 1	& 0	       & 5	        & 5 \\
Tool/Method Ref. Variations & 0	& 0	    & 0	            & 0	    & 0	      & 0	& 0	        & 0	         & 0	& 11	   & 0	        & 11 \\
Mapping Variations               & 0	& 4	    & 4	            & 0	    & 0	      & 0	& 0	        & 0	         & 0	& 0	       & 1	        & 1 \\
Appendix Variations              & 0	& 21	& 21	        & 0	    & 0	      & 0	& 0	        & 0	         & 0	& 0	       & 0	        & 0 \\
\hline
                                 & 34	& 133	& \textbf{167}  & 17	& 0	      & \textbf{17}	& 52 & 20 & \textbf{72} & 58	   & 26	        & \textbf{84} \\
\hline
\end{tabular}
\end{table}
\begin{figure}[ht]
\begin{minipage}[b]{0.48\textwidth} \centering
	\centering
	\includegraphics[width=\columnwidth]{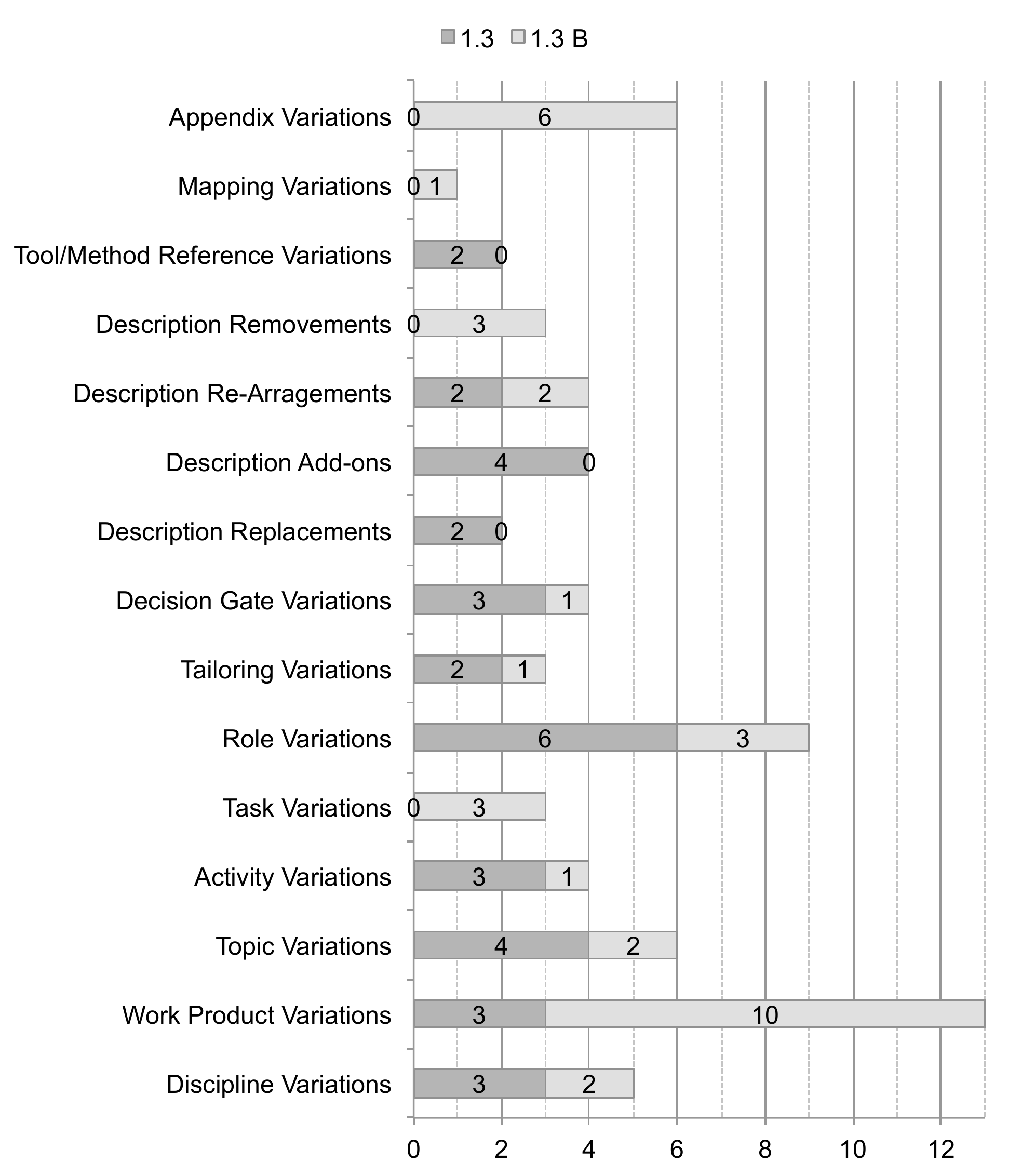}
	\caption{Defined variability operation types per metamodel version.}
	\label{fig:VarOpsComplete}
\end{minipage}
\quad
\begin{minipage}[b]{0.49\textwidth} \centering
	\centering
	\includegraphics[width=\columnwidth]{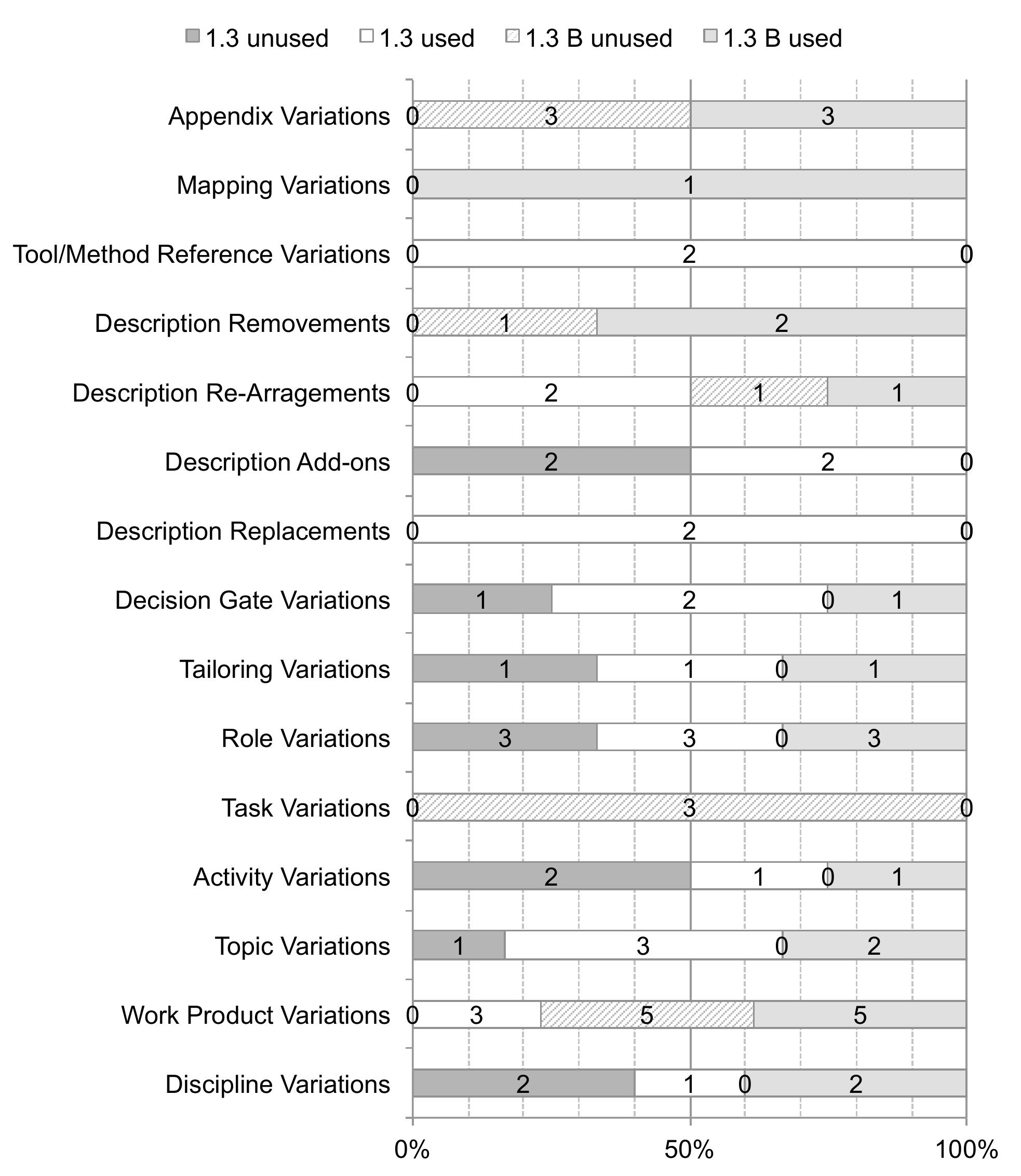}
	\caption{Used/unused variability operation types per metamodel version.}
	\label{fig:VarOpsUsedByMM}
\end{minipage}
\end{figure}
Figure~\ref{fig:VarOpsUsedByMM} quantifies the use within the operation groups and per metamodel version (cf.~Table~\ref{tab:VarOpsByTypeMMRelease}). An operation type is in the set of used operations if there is at least one exemplar in any of the studied variants.
Figure~\ref{fig:VarOpsUsedByMM} shows which metamodel defines how many operation types (per operation group) and how many of them are used in the  variants (overall count). 

Table~\ref{tab:VarOpsByTypeMMRelease} gives the more detailed perspective based on the exemplars per operation group. The table does not list V-Modell~XT variant ``D'' (Figure~\ref{fig:VMProcessLineOverview}), as this variant does not contain any variability operation exemplars, i.e., this variant uses SPrL features, but does not use the variability operation instrument (cf.\ Section~\ref{sec:RQ3}).

\paragraph{Unused Variability Operations}
Table~\ref{tab:UnusedVarOps} lists the defined, but unused operation types. The metamodel ``1.3'' defines 34 variability operation types of which 12 remain unused (35.3\%). The metamodel ``1.3B'' introduces 35 new variability operation types; 13 thereof remain unused (37.1\%). In summary, across all metamodel versions, 25 out of 69 (36.2\%) operation types are not used. 

\paragraph{Most Frequently Used Variability Operations}
Several operation types are frequently used across several process variants. Table~\ref{tab:MFUVarOps} lists the top-10 of the most frequently used operation types, including their number of exemplars and the metamodel that defines a particular operation type. The data shows that most of the frequently used operations modify text fragments of the process description (e.g., \emph{ReplaceSectionText}, \emph{ArrangeSection}) or alter the role model (e.g., \emph{ReplaceRoleDescription}, \emph{RenameRole}). Furthermore, the operations listed in Table~\ref{tab:MFUVarOps} contribute 224 of the 340 investigated operation exemplars, i.e., the top-10 operation types contribute 65.9\% of all found operation exemplars.
\begin{table}[t]
\begin{minipage}[t]{0.48\textwidth} \centering
	\centering

\renewcommand{\arraystretch}{1.15}
\footnotesize

\caption{Unused variability operation types.} 
\label{tab:UnusedVarOps}
\centering
\begin{tabular}{ll}
\hline
	\textbf{Operation Type} & \textbf{MM}\\
\hline
\hline
	AddDisciplineDescriptionPrefix & 1.3 \\
	AddDisciplineDescriptionPostfix & 1.3 \\
	DeleteWorkProduct & 1.3B \\
	ChangeWorkProduktDiscipline & 1.3B \\
	RenameCreatingDependency & 1.3B \\
	RenameTailoringDependency & 1.3B \\
	ReplaceTailoringDependencyDescription & 1.3B \\
	ArrangeSubTopic & 1.3 \\
	AddActivityDescriptionPrefix & 1.3 \\
	AddActivityDescriptionPostfix & 1.3 \\
	RemoveTask & 1.3B \\
	RenameTask & 1.3B \\
	ReplaceTaskDescription & 1.3B \\
	RemoveResponsibility & 1.3 \\
	AddRoleDescriptionPrefix & 1.3 \\
	RefineRole & 1.3 \\
	AddProcessModule & 1.3 \\
	AddDecisionGateDescriptionPrefix & 1.3 \\
	AddChapterTextPrefix & 1.3 \\
	AddSectionTextPrefix & 1.3 \\
	ChangeSectionNumber & 1.3B \\
	RemoveChapter & 1.3B \\
	RemoveGlossaryItem & 1.3B \\
	ReplaceGlossaryItemDescription & 1.3B \\
	RemoveAbbreviation & 1.3B \\
\hline
\end{tabular}

\end{minipage}
\quad
\begin{minipage}[t]{0.49\textwidth} \centering
	\centering
	
\renewcommand{\arraystretch}{1.15}
\footnotesize 

\caption{Most frequently used variability operations.} 
\label{tab:MFUVarOps}
\centering
\begin{tabular}{lll}
\hline
	\textbf{Operation Typ}e & \textbf{QTY} & \textbf{MM} \\
\hline
\hline
	ReplaceSectionText        & 46 & 1.3 \\
	ChangeRoleClass           & 36 & 1.3B \\
	ReplaceRoleDescription    & 34 & 1.3B \\
	RenameRole                & 22 & 1.3 \\
	RemoveLiteratureReference & 19 & 1.3B \\
	RemoveTopicAssignment     & 16 & 1.3B \\
	ChangeResponsibility      & 16 & 1.3 \\
	RemoveSupportingRole      & 12 & 1.3B \\
	ArrangeSection            & 12 & 1.3 \\
	ChangeDisciplineNumber    & 11 & 1.3B \\
\hline
\end{tabular}

\end{minipage}
\end{table}

\paragraph{Interpretation}
As 25 operation types remain unused, one may conclude that about one third of the variability operations seems to be dispensable. The reason for the existence of such operations is mainly for process language completeness. For instance, the metamodel defines a pair of \emph{Rename*} and \emph{Replace*} operations for each of the process dependency types. The definition of these operations was a design decision during the development of the metamodel ``1.3B''. Since the V-Modell~XT is designed as a generic framework, we cannot judge yet the relevance of the unused operation types, as future process variants may use them. 

Our findings also show that the most frequently used operations address the customization of description texts, e.g., \emph{ReplaceSectionText} or \emph{ReplaceRoleDescription}. For instance, in one process variant, the entire introduction to the process model was ``rewritten'' to reflect the special requirements of the current context, e.g., new procedures for handling quality gates, support to conduct the tailoring for the project context, improved integration of the process variant with the company-wide quality management, and integration of agile practices and support tools (use of \emph{ReplaceSectionText}\footnote{Note: An operation exemplar may address small and local modifications as well as extensive modifications that impact the whole process. Furthermore, a particular process model element can be addressed by multiple variability operations.}). Another process variant frequently used the operation \emph{ReplaceRoleDescription} to replace the generic role profiles by those established in the particular organization to align the customized process with the actual way of working. In the context of SPrLs, we interpret the frequent use as a standard use case in which a generic process description and a generic role model are refined for a particular process variant. Apart from just replacing text elements, the process's structure is also subject to change. For example, the operations \emph{ChangeResponsibility} and \emph{ArrangeSection} modify the process structure by modifying the responsibility for work products (work products are assigned to new/other roles) and, respectively, restructuring the process documentation (reordering existing chapters or injecting new ones).
Furthermore, we found variability operations also applied to compensate metamodel evolution.
For instance, the operation type \emph{Change\-RoleClass}, which is on the second rank (Table~\ref{tab:MFUVarOps}, 36 exemplars in two process variants), does not change any description text, but modifies the structure of role definitions in the process model in response to a metamodel evolution. In particular, this operation adds further metadata to ``legacy'' role elements that is required in newer versions of the metamodel. A variability operation is then used to reuse and update those role model elements without the need to change the reference model.  

In summary, beside content-related variability operations, we also found variability operations modifying the structure of process assets, e.g., to enable for backward-compatibility. However, the analysis of variant ``D'' (no operation exemplars) also shows that variability operations are only one instrument among others and, thus, SPrLs can also be created and managed using other mechanisms.

\subsection{RQ~3: Further Strategies to Provide Variability}
\label{sec:RQ3}
The third research question aims at studying whether there are situations in which variability is required, but not implemented using the variability operation instrument. We found two process variants in which variability operations were not applied, but where other strategies to realize variability were used. Table~\ref{tab:VMXT:NonVarStrategies} provides a description of these variability strategies. 
\begin{table}[h!t]
\renewcommand{\arraystretch}{1.3}
\footnotesize

\caption{Variability strategies not using the variability operation instrument.} 
\label{tab:VMXT:NonVarStrategies}
\centering
\begin{tabular}{p{0.1\textwidth}p{0.84\textwidth}}
\hline
	\textbf{Strategy} & \textbf{Description} \\
\hline
	Masking & Masking is, basically, no variability operation, although it could be considered as such. \emph{Pre-tailoring} is a built-in mechanism that allows for a coarse-grained modification of configuration containers (e.g., to remove whole sub-processes). Furthermore, to construct a process variant, SPrLs also allow for adding (completely) new process assets. The combination of pre-tailoring and adding new content can be used in a strategy called ``masking'' that allows for \emph{replacing} whole sub-processes. 
In the studied V-Modell~XT variants, we found two cases in which masking was used to realize variability (top-level configurations, project type variants, were removed from the reference process and substituted by similar ones). 
\newline
\emph{Rationale:} In several project type variants, existing process modules should be replaced by equivalent, but customized ones. However, no variability operation was defined to perform this replacement. Moreover, it turned out that a variability operation could not be implemented, as operation exemplars refer by \emph{id} to the process assets to be modified. The used metamodel, however, did not provide these attributes for the specific model element referring the process modules in the process configuration. That is, the missing variability operation caused by a gap in the metamodel was ``faked'' using masking and, thus, is not traceable during an automatic compliance check anymore. \\
	New Sub-\newline Processes & The analysis of variant ``D'' showed a setting in which a variant was derived without using variability operations. As mentioned before, SPrLs also allow for deriving a process variant by adding new content. 
Variant ``D'' is an example: The reference model was just taken and \emph{extended} by new content, e.g., new process modules, new roles, and new project type variants assembling the new process modules and such from the reference model. The new content showed no need for the use of variability operations as, for instance, no re-naming or text replacements were necessary. \\
\hline
\end{tabular}
\end{table}

In the \emph{new sub-processes} strategy, new (sub-)processes were introduced. This strategy does not require the use of variability operations and, essentially, realizes `true' extensions. However, both instruments can also be combined, e.g. in variant ``B'' and variant ``C'', new sub-processes were introduced, and variability operations were also used. Furthermore, in the variant ``Bund'', we found the \emph{masking} strategy, which was used to compensate a technical gap in the metamodel. In this strategy, several other operations were used to mimic missing variability operations.

\paragraph{Interpretation}
Variability operations are a meaningful tool. However, there are settings in which variability operations seem unnecessary. For instance, if a process variant mainly comprises added process assets while strictly adhering to a given reference process, variability operations may not be necessary (e.g., for the V-Modell~XT variant ``D''). Furthermore, there are settings in which variability operations cannot be defined due to metamodel-related gaps, but would be beneficial (e.g., in the V-Modell~XT variant ``Bund''). Settings using the masking strategy point to candidates for future metamodel improvements, such as adding \emph{id} tags and defining appropriate new variability operations.

\subsection{Validity of the Results}
\label{sec:EvaluationOfValidity}
We evaluate our findings and critically review our study regarding the threats to validity. 
The \emph{internal validity} could be threatened by a bias toward the variant construction process, because two of the authors are also the developers of the metamodels (and partially the processes). We minimized this threat by relying on an analysis tool, which was applied to all variants, and by calling in a third independent researcher. The \emph{external validity} is threatened as we have little knowledge to what extent we can generalize our results, e.g., to other SPrLs, as there are no similar cases available so far. As our study provides the first analysis of variability in this context, a generalization of the findings is, so far, not our intention. Instead, we are interested in analyzing the feasibility of variability operations based on given case, and to prepare future research on SPrLs.

\section{Conclusion}
\label{sec:Conc}
\label{sec:SummaryOfConc}
Our main goal was to analyze the feasibility of the variability operation instrument to support the systematic development of software process lines, and to develop an initial catalog of practically used variability operations. To this end, we studied the V-Modell~XT SPrL, and analyzed the reference process and 5 variants using the built-in SPrL features. So far, we found two metamodel versions defining variability operations: the metamodel of the reference process defines 34 variability operation types, and the enhanced metamodel of the V-Modell~XT Bund adds 35 more types. In summary, we found 69 variability operation types, which allow process engineers to declaratively modify process content (e.g., by providing new text snippets) and to modify the structure of a process (e.g., by changing responsibilities, removing references, and modifying the tailoring behavior). Furthermore, we investigated which variability operations were applied in practice, which allows us to rate the feasibility of this instrument. Among the 69 identified operation types, we found that 25 where defined, but remained unused. Unused operation types were either defined during the initial development of this instrument (based on experiences or to allow for a constrictive compliance), or the operations were defined during metamodel improvements (mainly to improve the completeness of the process language). 
Our findings also show that the variability operation instrument also serves metamodel evolution, which inherently happens in long-term SPrL development. In particular, we found variability operations that allow for structurally modifying `legacy' process assets for reuse in newer versions of a process.
Finally, we found settings in which variability operations were not used, as they were either unnecessary or missing. If variability operations were missing, we found a work around used to mimic missing operation types and, thus, we could also identify metamodel improvement candidates.

In summary, we found the variability operation instrument sufficient to support process engineers in constructing a (new) process variant from an SPrL. However, variability operations are only one instrument among others and they can be combined with other instruments. We also showed the difficulty to define meaningful variability operations as we found a number of variability operations defined, but unused. Still, we could find for all these operations a rationale for why they are part of the model whereby further evaluation needs to remain in scope of future investigations.

Practitioners can benefit from our findings. As variability operations are a means to declaratively define modifications of a reference process, this concept offers payoffs in domains in which regularized processes must be applied, e.g., in medicine, automotive, and in avionics. A company-specific process can declare the modifications regarding the reference process using variability operations that can be easily tracked and, thus, support audits and assessments. Furthermore, using variability operations helps to reduce process maintenance cost. In the optimal case, the company-specific process can be updated by replacing the old reference model by a newer version and by executing the merge procedure again. Furthermore, due to the separation of the reference- and the extension model, companies can establish a process life cycle management independent from the reference model. That is, as a company is in full control of its own variant, it can follow its own maintenance and improvement cycles, and, furthermore, can decide which of potentially refreshed reference contents is adopted for the company's variant(s).

\paragraph{Relation to Existing Evidence}
\label{sec:RelationToExistingEvidence}
In \cite{AKM+08,AKM+09,OM09,OMR09,OS07}, initial research was done in the area of evolution-driven variability analysis. However, these contributions aim at identifying variations in given models. Our research is focussed on a constructive approach that supports variability by design. Mart\'{i}nez-Ruiz et al.\ \cite{Martinez-Ruiz:2012fk} conducted a study in which they investigated the constructors used in tailoring, revealing that current tailoring constructors do not meet industry requirements. They argue for instruments allowing for variability and consistency at the same time. Carvalho et al.\ \cite{Carvalho:2014zr} provide a systematic literature review of SPrLs and their implementation in practice, and conclude that SPrL concepts still have to be considered immature due to a lack of reported empirical evidence. 

Due to its dissemination, SPEM is a candidate for sophisticated variability concepts, and as we found in \cite{kms2013}, a number of improvement proposals are already available. SPEM defines a set of basic variability operations (e.g., extends or replaces), which is the basis for several SPrL-related improvement proposals (see, e.g.,~\cite{al.:1900bh,Martinez-Ruiz:2013:PVM:2486046.2486056, Martinez-Ruiz-T.:2011uq,Oliveira:1900cr}). However, no case study in the context of SPEM is yet available presenting concrete practical experiences.  

\paragraph{Limitations}
\label{sec:Limitations}
The major limitation is that our study is based on the V-Modell~XT only. However, it is the only process framework that provides this explicit support to create SPrLs. Furthermore, directly comparing the V-Modell~XT framework and SPEM, significant differences regarding the notion of process variability as well as process tailoring become evident. While process customization (at the company level) and process tailoring (at the project level) are clearly separated processes in the V-Modell~XT, SPEM does not provide such a clear differentiation. Therefore, as the concepts as well as the processes differ, transfer and generalization of the findings have to be prepared carefully. Furthermore, the V-Modell~XT provides a rich portfolio of instruments to create process variants. In this article, we focused on the variability operation instrument, and we barely scratched the surface regarding other instruments (e.g., Section~\ref{sec:RQ3}).

\paragraph{Future Work}
\label{sec:FutureWork}
As our investigation is based on a snapshot of the V-Modell~XT process line, which is based on the version 1.3 of the reference model and all related variants, the study at hand needs to be repeated for future versions of the reference model and its variants. A repeated analysis allows for analyzing the evolution of the instrument over time, e.g., regarding the question whether there are new variability operations (e.g., addressing the gaps discussed in Section~\ref{sec:RQ3}), or whether unused variability operations are removed or used in future versions. Such an analysis can be accompanied by an in-depth analysis investigating which elements are modified in the different variants and if there are variability pattern across the different variants, which can be used to derive particular improvement requirements for the reference model. 

As a second step, independent research is necessary to analyze the transfer options to other frameworks. Variability operations are a meaningful instrument to support process variability, however, as we already discussed in \cite{kk2013a} and as also mentioned in \cite{Martinez-Ruiz:2012fk}, there is a gap in process frameworks regarding the capability to model flexible processes. This gap needs to be closed and, thus, it needs to be further investigated whether variability operations can contribute.

\bibliographystyle{abbrv}
\bibliography{vmlines}

\end{document}